\documentclass[reprint, amsmath, amssymb, aps, pra, floatfix]
{revtex4-2}

\usepackage{graphicx}
\usepackage{dcolumn}
\usepackage{bm}
\usepackage{float}
\usepackage{xcolor}
\usepackage{physics}
\usepackage{adjustbox}
\usepackage{multirow}
\usepackage{array} 
\usepackage{nicematrix}
\usepackage{enumitem}
\usepackage[normalem]{ulem}
\usepackage[hidelinks]{hyperref}

\newlength{\mytablinewidth}
\newcolumntype{V}{!{\vrule width \mytablinewidth}}

\newcommand{\rbcs}{\ensuremath{\mathrm{^{87}Rb^{133}Cs}}}

\newcommand{\xssigma}{\ensuremath{\mathrm{X}\,^{1}\Sigma^{+}}}
\newcommand{\assigma}{\ensuremath{\mathrm{A}\,^{1}\Sigma^{+}}}
\newcommand{\btpi}{\ensuremath{\mathrm{b}\,^{3}\Pi_{0}}}

\newcommand{\Asbt}{\ensuremath{\mathrm{A}\,^{1}\Sigma^{+} - \mathrm{b}\,^{3}\Pi_{0}}}

\newcommand{\fig}[1]{Fig.~\ref{#1}}
\newcommand{\eq}[1]{Eq.~(\ref{#1})}
\newcommand{\tab}[1]{Table~\ref{#1}}

\newcommand{\tj}[6]{ \begin{pmatrix}
	#1 & #2 & #3 \\
	#4 & #5 & #6 
   \end{pmatrix}}

\begin{document}

\date{\today}
\title{Hyperfine-resolved optical spectroscopy of ultracold $^{87}$Rb$^{133}$Cs molecules: \\ the \btpi\, metastable state}

\author{Arpita~Das}
\email{arpita.das@durham.ac.uk}
\affiliation{Department of Physics and Joint Quantum Centre (JQC) Durham-Newcastle, Durham University, Durham DH1~3LE, United Kingdom.}
\author{Albert~Li~Tao}
\affiliation{Department of Physics and Joint Quantum Centre (JQC) Durham-Newcastle, Durham University, Durham DH1~3LE, United Kingdom.}
\author{Luke~M.~Fernley}
\altaffiliation[Present address: ]{National Quantum Computing Centre, Didcot OX11 0QX, United Kingdom}
\affiliation{Department of Physics and Joint Quantum Centre (JQC) Durham-Newcastle, Durham University, Durham DH1~3LE, United Kingdom.}
\author{Fritz~von~Gierke}
\altaffiliation[Present address: ]{Institut für Quantenoptik, Leibniz Universität Hannover, 30167 Hannover, Germany}
\affiliation{Department of Physics and Joint Quantum Centre (JQC) Durham-Newcastle, Durham University, Durham DH1~3LE, United Kingdom.}
\author{Philip~D.~Gregory}
\affiliation{Department of Physics and Joint Quantum Centre (JQC) Durham-Newcastle, Durham University, Durham DH1~3LE, United Kingdom.}
\author{Jeremy~M.~Hutson}
\affiliation{Department of Chemistry and Joint Quantum Centre (JQC) Durham-Newcastle, Durham University, Durham DH1~3LE, United Kingdom.}
\author{Romain~Vexiau}
\affiliation{Universit\'{e} Paris-Saclay, CNRS, Laboratoire Aim\'{e} Cotton, Orsay, 91400, France.}
\author{Olivier~Dulieu}
\affiliation{Universit\'{e} Paris-Saclay, CNRS, Laboratoire Aim\'{e} Cotton, Orsay, 91400, France.}
\author{Simon~L.~Cornish}
\email{s.l.cornish@durham.ac.uk}
\affiliation{Department of Physics and Joint Quantum Centre (JQC) Durham-Newcastle, Durham University, Durham DH1~3LE, United Kingdom.}

\begin{abstract}
Using an ultracold gas of \rbcs\, molecules, we perform hyperfine-resolved spectroscopy of transitions from the vibronic ground state to the lowest rovibrational states of the electronic state \btpi\, as a function of magnetic field. These transitions are spin forbidden, resulting in narrow linewidths, and feature near-diagonal Franck-Condon factors. We develop a model of the hyperfine and Zeeman structure that includes coupling between the $0^+$ and $0^-$ components of \btpi\,. We fit the spectra to obtain rotational and hyperfine coupling constants. We measure transition dipole moments associated with specific transitions by directly observing Rabi oscillations as a function of a resonant laser pulse duration. Using resonant $\pi$ pulses, we prepare molecules in the electronically excited state and directly measure the spontaneous emission rate.
\end{abstract}

\maketitle

\section{\label{sec:intro} Introduction}
\vspace{-0.25 cm}

Ultracold polar molecules have applications in the fields of quantum simulation and computation~\cite{Sawant2020, Kaufman2021,
Feng2022,Zhang2022,Cheng2023,Cohen2024,Cornish2024, Carroll2025, Miller2024}, quantum-state-controlled chemistry~\cite{Krems2008, Ospelkaus2010, Langen2024}, and precision measurement of fundamental constants~\cite{Klos2022, Kimball2023, DeMille2024, Huang2024}. Dipole-dipole interactions between molecules can be precisely engineered using microwaves or static electric fields resulting in detectable quantum entanglement over long range~\cite{Yan2013,Bao2023, Holland2023, Picard2025, Ruttley2025}. Molecules also possess rich internal structure due to the combination of electronic, vibration, rotation, and nuclear-spin degrees of freedom. This presents opportunities for experiments; for example, vibrational transitions allow precise clock measurements in the THz frequency domain~\cite{Leung2023}, while rotational and hyperfine states can be used to encode interacting~\cite{Bao2023, Holland2023, Picard2025, Ruttley2025} and storage~\cite{Park2017, Ni2018, Gregory2021} qubits. 

At present, the coldest and densest samples consist of bialkali molecules prepared by association from pre-cooled atomic mixtures~\cite{DeMarco2019,Schindewolf2022,Bigagli2023,Lin2023, Bigagli2024}. The RbCs molecule is the heaviest heteronuclear combination of two alkali atoms, and was the second polar molecule to be produced by associating atom pairs. This is a two-step process. Initially, weakly-bound molecules are produced by magnetoassociation~\cite{Takekoshi2012,Koppinger2014}. After formation, molecules are transferred to the \xssigma\, ground state using stimulated Raman adiabatic passage (STIRAP)~\cite{Takekoshi2014,Molony2014}. From here, electronically excited states can be accessed with transition frequencies greater than 261.5\,THz. The lowest electronically excited states are the interacting \Asbt\, and \btpi\, states, which are mixed by spin-orbit interactions~\cite{Fahs2002}; their potential curves are shown in Fig.~\ref{fig:RbCsPotential}. Current experiments exploit the mixing to achieve effective coupling between the triplet state produced by magnetoassociation and the singlet rovibronic ground state. 

Our group has recently demonstrated a magic-wavelength trap for \rbcs\, that allows second-scale coherences to be engineered between rotational states~\cite{Gregory2024,Ruttley2025,Hepworth2025}. This trap uses light that is carefully tuned between transitions to the lowest two vibrational levels of the \Asbt\, system, where it has predominantly \btpi\, character, to suppress the anisotropic components of the polarisability tensor~\cite{Guan2021}. These transitions have narrow linewidths because they are nominally forbidden. Knowledge of the energies and linewidths associated with these transitions is important for understanding the magic trap. 

Transitions to the lowest-energy levels of \btpi\, may be useful for other applications. In particular, they have highly diagonal Franck-Condon factors and may offer nearly closed optical cycles for direct laser cooling~\cite{Shuman2010} or absorption imaging~\cite{Wang2010} of the molecules. Moreover, they may allow implementation of schemes for two-photon collisional shielding~\cite{Karam2023} and non-destructive detection~\cite{Guan2020} of molecules. Spectroscopy of the lowest vibrational levels of \btpi\, has been performed for other bialkali molecules KRb~\cite{Kobayashi2014}, NaK~\cite{Bause2020}, NaRb~\cite{He2021}, and very recently LiK\,~\cite{Yang2025}, but not for RbCs.

In this article, we perform hyperfine-resolved spectroscopy of the lowest vibrational levels of the \btpi\, state of \rbcs\, (hereafter RbCs) by driving spin-forbidden \xssigma\, $\rightarrow$ \btpi\, transitions. We prepare molecules in the rovibrational and hyperfine ground state of the \xssigma\, state at a magnetic field of 181.6\,G by magnetoassociation followed by STIRAP. We then pulse on light to drive electronic transitions in the molecule. We observe loss of molecules from  \xssigma\, when the light is resonant with an \xssigma\, $\rightarrow$ \btpi\, transition. Our work focuses on the transitions $(v'' = 0, J'' = 0,\, 1) \rightarrow (v' = 0,\, 1, J' = 0,\, 1,\, 2)$ that occupy the region 261.6\,THz to 264.6\,THz, where $v'',\, J''$ and  $v',\, J'$ are the vibrational and rotational quantum numbers of the ground and excited state respectively. These are the transitions closest to the rotationally magic condition observed in~\cite{Gregory2024}.

\section{Background}

\subsection{The mixed electronic states \assigma\,--\,\btpi\, of RbCs}

The level structure of the mixed electronic states \assigma\,--\,\btpi\, that correlate with the dissociation limit Rb(5s)+Cs(6p) has been characterised for RbCs over a broad energy range by measurements of laser-induced fluorescence and Fourier transform spectroscopy in heat pipes~\cite{bergeman2003,Docenko2010,Kruzins2014}. The uncertainty in these measurements is estimated to be $\pm0.3$\,GHz, with measurements limited by a combination of spectrometer resolution and Doppler broadening. Spectroscopy using ultracold samples allows the structure to be resolved more precisely by suppressing Doppler broadening. Kerman~\emph{et al.}~\cite{Kerman2004} studied photoassociation lines in a dual-species magneto-optical trap that probes the long-range part of the potential. Debatin~\emph{et al.}~\cite{Debatin2011} performed hyperfine-resolved spectroscopy on ultracold molecules produced by magnetoassociation, accessing states in the range $304.6$ to $307.6$\,THz from the ground state with an accuracy of $\pm10$\,MHz, to identify an optimum intermediate level for STIRAP. However, no work has previously reported direct measurements of the lowest vibrational levels of $\mathrm{b}\,^3\Pi$ in RbCs. 

\begin{figure}[t!]
	\centering
	\includegraphics[width=0.48\textwidth]{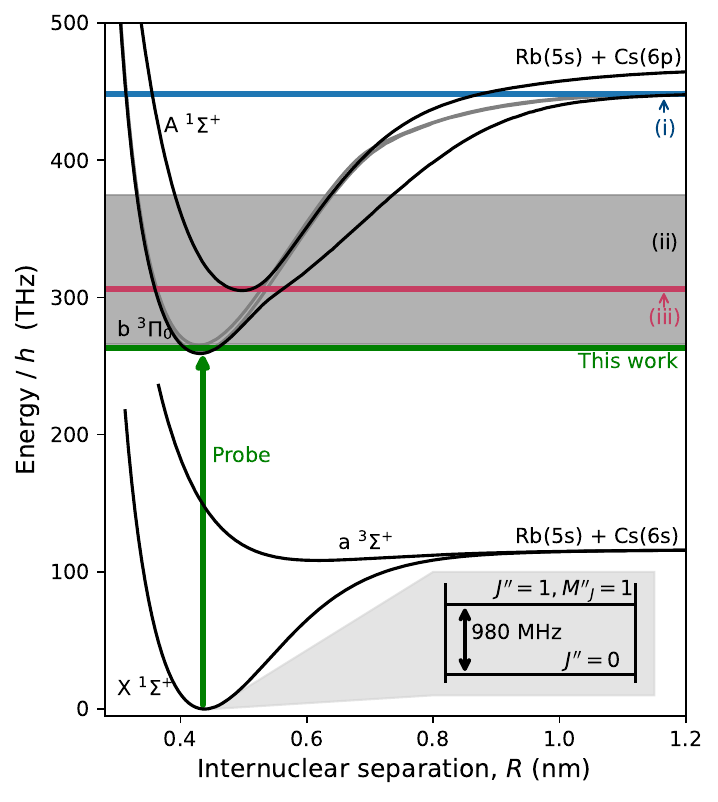}
	\caption{\label{fig:RbCsPotential} The relevant electronic potentials for RbCs. In the excited state, the potentials for the mixed $\mathrm{A}\,^1\Sigma$ and $\mathrm{b}\,^3\Pi_0$ states are shown by black lines, while those for $\mathrm{b}^3\Pi_1$ and $\mathrm{b}\,^3\Pi_2$ are shown in grey. The shaded areas indicate regions of the mixed states that have been investigated by (i)~photoassociation in magneto-optical traps~\cite{Kerman2004}, (ii)~Laser-induced fluorescence and Fourier transform spectroscopy in heat pipes~\cite{Docenko2010,Kruzins2014}, (iii)~absorption spectroscopy with ultracold molecules produced via magnetoassociation~\cite{Debatin2011}. The energy region studied in this work is shown in green, with the vertical green arrow labelled ‘probe’ indicating the \xssigma\, $\rightarrow$ \btpi\, transitions. The inset illustrates the rotational structure of the \xssigma,\, $v = 0$ ground state. 
    }
\end{figure}

Spin-orbit coupling with the \assigma\, state and the two other states B\,$^1\Pi$ and c\,$^3\Sigma^+$ that correlate with Rb(5s)+Cs(6p) splits the $\mathrm{b}\,^3\Pi$ state into separate $\mathrm{b}\,^3\Pi_0$, $\mathrm{b}\,^3\Pi_1$, $\mathrm{b}\,^3\Pi_2$ components, corresponding to projections $\Omega=0,1,2$ of the total electronic angular momentum onto the molecular axis. The coupling between \assigma\, and \btpi\, is strongest around the crossing of the two potentials at an internuclear separation of 0.53~nm, resulting in the avoided crossing shown in \fig{fig:RbCsPotential}. The lowest vibrational levels of the \btpi\, state are well below this crossing and so are only weakly mixed, with $<0.5\%$ singlet character. Nevertheless, this small mixing is sufficient to allow us to drive transitions from the \xssigma\, ground state. In addition, spin-orbit coupling with the $\mathrm{c}\,^3\Sigma^+$ state (well above \xssigma\, in energy) splits the \btpi\, state into two components, $\mathrm{b}\,^3\Pi_{0^+}$ and $\mathrm{b}\,^3\Pi_{0^-}$, where the sign indicates the parity of the electronic wave function with respect to the plane containing the molecular axis.

\subsection{Production and detection of RbCs molecules}
We create ultracold RbCs molecules by magnetoassociation from a thermal mixture of $5\times10^5$ $^{87}$Rb atoms and $3\times10^5$ $^{133}$Cs atoms. The atoms are prepared at a temperature of $\sim300$\,nK in a magnetically levitated crossed optical dipole trap of wavelength $\lambda=1550$\,nm, and occupy their internal ground states, ($f_\mathrm{Rb}=1$, $m_{f_{\mathrm{Rb}}}=1$) and ($f_\mathrm{Cs}=3$, $m_{f_{\mathrm{Cs}}}=3$)~\cite{McCarron2011}. To form molecules, we sweep the magnetic field down across an interspecies Feshbach resonance at 197.10\,G~\cite{Takekoshi2012}, following the scheme established in~\cite{Koppinger2014}. We purify the sample of molecules by removing any remaining atoms from the trap using the Stern-Gerlach effect~\cite{Herbig2003}. We then transfer the molecules to an unlevitated optical potential by first increasing the optical trap depth, which compresses the molecules and adiabatically heats the sample to a temperature of $1.5\,\mu$K. After this the magnetic field gradient is ramped off without changing the temperature of RbCs cloud. Finally, the dipole trap is switched off, leaving the molecules in free space, where they are transferred to the rovibrational ground state of \xssigma\, using STIRAP, with a typical efficiency of $\sim90\%$~\cite{Molony2014,Molony2016}. The STIRAP is performed at a magnetic field of 181.6\,G; except for measurements of Zeeman shifts, all our spectroscopy is performed at this field. 

To detect the molecules we reverse the STIRAP and dissociate; absorption imaging of the resulting atoms allows us to count the number of molecules that were present. This method detects only molecules that are in the specific hyperfine state in which they were initially prepared. There is no change in the observed molecule number for hold times in free space up to 1\,ms. After this time, the number decreases as the molecules fall out of the $\sim35\,\mu$m region addressed by the STIRAP beams.

We label the states in a given electronic and vibrational state by quantum numbers $(J, M_F)_\mathrm{G,E}$. Here $J$ is a pure rotational quantum number for the electronic ground state, but includes contributions from orbital and electron spin angular momenta for the excited state. $M_F=M_J+m_\mathrm{Rb}+m_\mathrm{Cs}$ is the projection of the total angular momentum, including the projections $m_\mathrm{Rb}$ and $m_\mathrm{Cs}$ of the nuclear spins $I_\mathrm{Rb}$ and $I_\mathrm{Cs}$. $J$ is a nearly good quantum number because the hyperfine and Zeeman effects are small compared to the rotational spacing, but the total angular momentum $F =J + I_\mathrm{Rb} + I_\mathrm{Cs}$ is not conserved in the presence of a magnetic field. The subscript denotes whether the state is in the electronic ground state (G) or the excited state (E). In this work, the ground-state molecules are prepared in either $(0,5)_\mathrm{G}$ or $(1,6)_\mathrm{G}$. Both these states are spin-stretched so have well-defined quantum numbers in the uncoupled basis $(J, M_J, m_\mathrm{Rb}, m_\mathrm{Cs})$:
\begin{equation*}
(0,5)_\mathrm{G} \equiv (0, 0, 3/2, 7/2);
\end{equation*}
\begin{equation*}
(1,6)_\mathrm{G} \equiv (1, 1, 3/2, 7/2).
\end{equation*}
For other states, $M_F$ is conserved but $M_J$, $m_\mathrm{Rb}$ and $m_\mathrm{Cs}$ are not individually good quantum numbers.
The state $(0,5)_\mathrm{G}$ is the one that is populated by the STIRAP. For spectroscopy of molecules in the initial state $(1,6)_{\mathrm{G}}$, we use two microwave $\pi$-pulses~\cite{Gregory2016}: one to transfer the molecules from $(0,5)_{\mathrm{G}}$ to $(1,6)_{\mathrm{G}}$ and one to return the molecules to $(0,5)_{\mathrm{G}}$ after exposure to the probe light.

\subsection{{\label{sec:transitions}}Driving $\mathrm{X}-\mathrm{b}$ transitions}
We expect electronic transitions from \xssigma\, to obey the electric dipole selection rules, $\Delta J = \pm 1$, and $\Delta M_J = 0, \pm1$, where the latter depends on the polarisation of the light. We also expect that the allowed transitions couple only state components that preserve the nuclear spin projections, such that the selection rule on $\Delta M_J$ effectively implies $\Delta M_F=0, \pm1$. The strength of the transitions observed will therefore depend on the transition dipole moment for the electronic transition, the selection rules for $J$ and $M_J$, and the nuclear spin composition of the specific hyperfine states being addressed. 

We use an external-cavity diode laser (Toptica DL pro) to generate the probe light (1133 to 1147\,nm). To ensure long-term frequency stability, we couple the laser to an optical cavity with an ultra-low-expansion glass spacer (Stable Laser Systems). The finesse of the cavity in this wavelength range is approximately $37,000$ and the free spectral range (FSR) is $1,496.755\,172\,9(9)$ MHz. We offset-lock the laser frequency to a cavity mode using the Pound-Drever-Hall (PDH) technique~\cite{Gregory2015}. We use a fast feedback loop (Toptica FALC), which helps to reduce the laser linewidth. We estimate that the laser linewidth is $4.7(1)$\,kHz from the deviation of the error signal when locked. The probe light is delivered to the main experiment using a polarisation-maintaining optical fibre. We deliver up to 1.4\,mW of light to the molecules with a beam waist of 1.09(5)\,mm. This waist size is much larger than the size of the sample ($\sim10\,\mu$m) which ensures that all molecules are illuminated uniformly. 

We perform spectroscopy in two geometric configurations. First, we have the light propagating perpendicular to the magnetic field that sets the quantisation axis. In this case, the light is linearly polarised either parallel or perpendicular to the quantisation axis to drive either $\pi$ or $\sigma^{\pm}$ transitions, respectively. In the second configuration, the light propagates at an angle of $\sim10^{\circ}$ to the quantisation axis and is circularly polarised to drive either $\sigma^+$ or $\sigma^-$ transitions preferentially. 

Our experimental procedure is as follows. We prepare the molecules in either $(0,5)_{\mathrm{G}}$ or $(1,6)_{\mathrm{G}}$ at $181.6$\,G and expose them to a $500\,\mu$s pulse of probe light before detecting the number remaining. We repeat this sequence, each time changing the laser frequency to find the excited-state resonances, which appear as loss in the number of molecules as we drive the transition to $(J',M'_F)_{\mathrm{E}}$. To begin, we perform an initial search for transitions using the highest probe power available. This yields power-broadened spectra with loss features that are saturated and flat-bottomed. For each loss feature observed at high power, we scan the frequency over the feature using progressively lower probe powers in order to desaturate the transition, typically aiming for a loss feature with depth $\sim 50$\% to 80\% of the initial molecule number. Each feature is therefore observed using a different probe power. We fit each loss feature with a Gaussian to extract the centre frequency.

\section{\label{sec:result} Results}

\subsection{{\label{sec:vib_structure}}Vibrational structure of \btpi\,}

\begin{figure}[t!]
    \centering
    \includegraphics[width=0.5\textwidth]{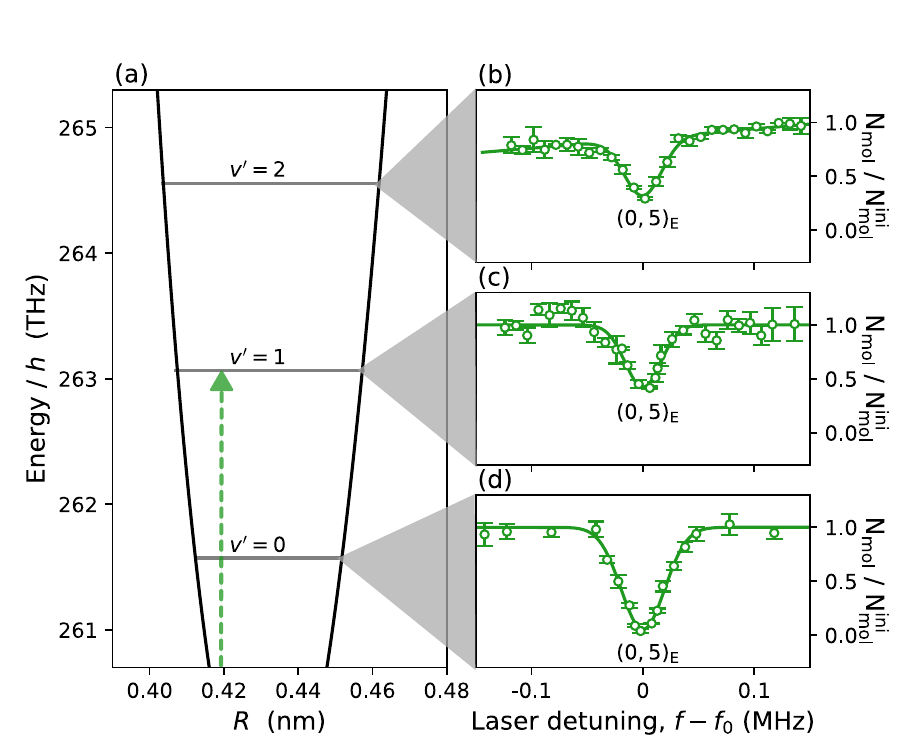}
    \caption{\label{fig:vib_spectra}~Characterisation of the vibrational structure. The \btpi\, electronic potential with the energies of $v' = 0, \,1, \, \textrm{and} \,2$ shown on the left. The green arrow indicates the probe laser driving transitions from $(1,6)_\mathrm{G}$ to the $(0,5)_\mathrm{E}$ state in each vibrational level. Spectroscopy on each of these transitions performed at 181.6\,G is shown on the right, with the laser detuning ($f-f_0$) plotted relative to the centre frequency $f_0$ of each of the observed transitions. We attribute the sloped background in the spectra for $v' = 2$ to nearby hyperfine structure combined with imperfect polarisation of the probe laser. In the analysis of this measurement only, we fit the results with a linearly varying background.}
\end{figure}

To study the vibrational structure of the \btpi\, state, we use light polarised perpendicular to the quantisation axis to drive transitions from $(1,6)_{\mathrm{G}}$ to $v' = 0, 1, 2,\, J' = 0$. In each case only one transition is accessible: the $\sigma^-$ transition to $(0,5)_{\mathrm{E}}$. The results are shown in \fig{fig:vib_spectra}. 
The absolute transition frequencies for each transition are shown in \tab{tab:vib_structure}, where the uncertainties ($\pm60$\,MHz) come from the wavemeter (Bristol 621A). The difference in transition frequencies is determined much more precisely ($\pm 2$\,kHz) by using the modes of the reference cavity as a frequency ruler. We compare the absolute transition frequencies we measure with those calculated from the molecular potential curves of refs.~\cite{Docenko2010, Docenko2011, Rakic2016, Vexiau2017}.

Anharmonicity of the potential curves leads to $\Delta E_{v'_{01}} > \Delta E_{v'_{12}}$, where $\Delta E_{v'_{01}}$ and $\Delta E_{v'_{12}}$ are energy differences between  excited vibrational levels $v ' = 0$ and $v' = 1$ and between $v' = 1$ and $v' =2$ respectively. The energies $E_v$ of the states near the bottom of the potential are approximately
\begin{equation}\label{eq:vib-energy}
    E_{v} = h \nu_\textrm{e}\left[\left(v+\textstyle{\frac{1}{2}}\right)-x_\textrm{e}\left(v+\textstyle{\frac{1}{2}}\right)^2\right],
\end{equation}
where $\nu_\textrm{e}$ is the harmonic-oscillator frequency. The second term in \eq{eq:vib-energy} gives the effect of the anharmonicity, characterised by $x_\textrm{e}$, which is a small unit-less number. Fitting the observed vibrational structure with \eq{eq:vib-energy}, we find $\nu_\textrm{e} =1.497\,603\,712(7)$~THz $ \equiv 49.9546827(2)$~cm$^{-1}$ and $x_\textrm{e}=1.275\,851(2)\times10^{-3}$. We have previously measured the binding energy of the rovibrational and hyperfine ground state of the \xssigma\, potential using an optical frequency comb~\cite{Molony2016a}. Using this result and the measurement reported here, we estimate the binding energy of the rovibrational ground state of the $\mathrm{b}\,^3\Pi_{0^+}$ state, with respect to the asymptote $\mathrm{Rb}(5^{2}\mathrm{S}_{1/2})+\mathrm{Cs}(6^{2}\mathrm{P}_{3/2})$, to be $D_{0} = h \times 205.17133(6)$~THz =\ $hc \times 6843.779(2)$~cm$^{-1}$, where the uncertainty is dominated by the precision of our wavemeter.

\begin{table}[b!]
\centering
\caption{\label{tab:vib_structure}
Results of spectroscopy to investigate the vibrational structure near the bottom of the \btpi\, potential. Absolute and relative frequencies for the transitions from $(1,6)_\mathrm{G}$ to $(0,5)_\mathrm{E}$ in $v' =0,\, 1,\, 2$, extracted from the spectroscopy shown in \fig{fig:vib_spectra}.}
    \begin{ruledtabular}
    \begin{tabular}{cccc}
    \multicolumn{1}{c}{$v'$} & \multicolumn{2}{c}{Transition frequency (THz)} &  \multicolumn{1}{c}{$\Delta E_{v_{ij}}/h$ (GHz)} \\
    &Theory~\cite{Svetlana2021}&This work&         \\
    \hline
	$0$& $261.531$ & $261.56987(6)$  &  \\ 
	$1$& $263.034$ &$263.06363(6)$  & $\Delta E_{v_{01}}=1493.782274(2)$ \\ 
	$2$& $264.531$ & $264.55358(6)$ & $\Delta E_{v_{12}}=1489.960836(2)$  \\ 
    \end{tabular}   
    \end{ruledtabular}   
\end{table}

\begin{figure*}[t!]
    \includegraphics[width=0.9\textwidth]{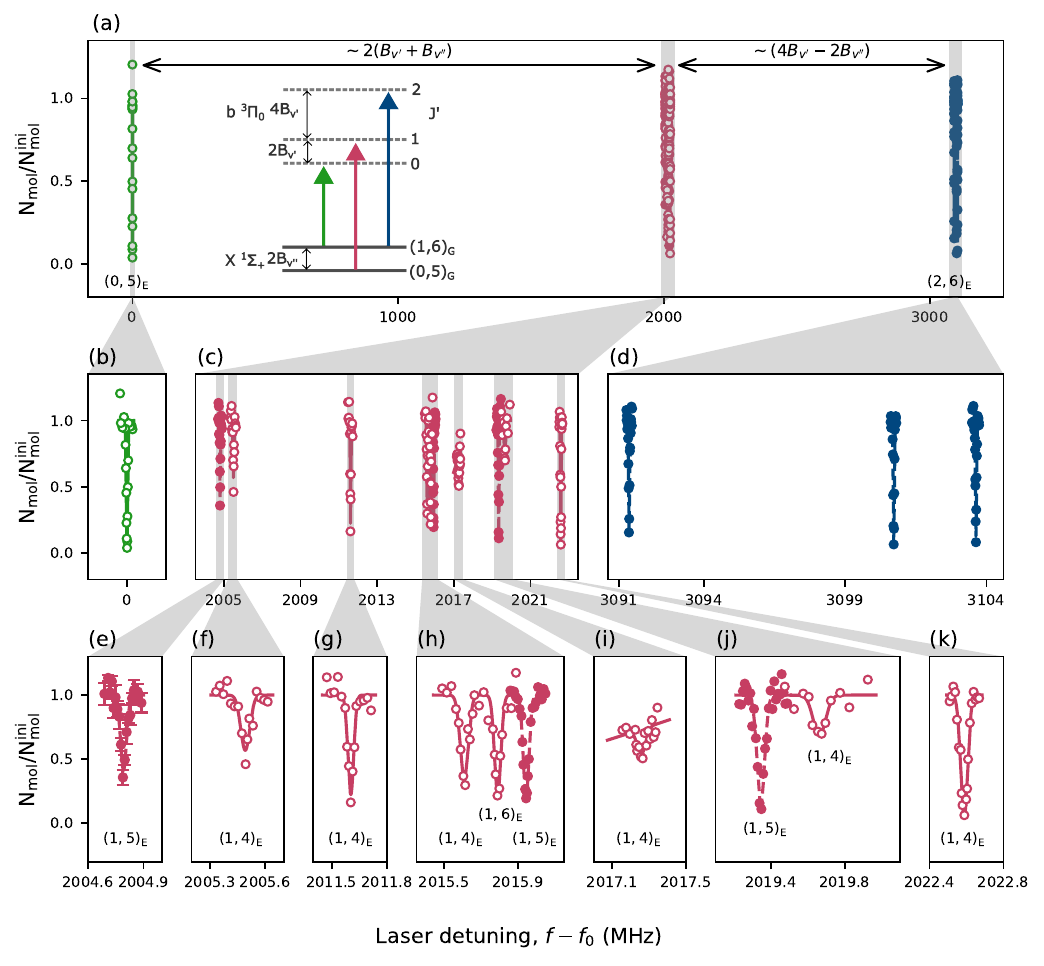}
    \caption{\label{fig:all_transitions} An overview of rotational and hyperfine structures of the transitions at 181.6\,G for $v' = 0$. (a) The coarse rotational structure. The inset shows the level diagram of the transitions we drive. The colours of the vertical arrows match the corresponding data points. (b), (c) and (d) show the observed hyperfine structure. (b) The $\sigma^{-}$ transition from $(1,6)_\mathrm{G}$ to $(0,5)_\mathrm{E}$ (also shown in \fig{fig:vib_spectra}(d)). (c) $\pi$ and $\sigma^{\pm}$ transitions from the $(0,5)_\mathrm{G}$ state to $(1,M'_F)_\mathrm{E}$ states. (d) $\pi$ transitions from the $(1,6)_\mathrm{G}$ state to $(2,6)_\mathrm{E}$ states. (e)\,--\,(k) show the zoomed-view of all transitions we observed in (c). All transitions in (e)\,--\,(k) are marked with the character of the hyperfine state $(J',M'_F)_\mathrm{E}$. The x axis scale of (e)\,--\,(k) are the same. For a clearer view of the transition profiles, we have shown the error bars only on (e), while all other data have similar error bars.}
\end{figure*}

\subsection{{\label{sec:rot_hf_structure}}Rotational and hyperfine structure}

In \fig{fig:all_transitions} we present spectroscopy of the rotational and hyperfine structure associated with the $v' = 0$ level of the \btpi\, state at 181.6\,G. \fig{fig:all_transitions}(a) shows the rotational structure of the transitions. The green feature indicates the transition $J'' = 1 \rightarrow J' = 0$, while the red and blue features indicate the transitions $J'' = 0 \rightarrow J' = 1$ and $J'' = 1 \rightarrow J' = 2$, respectively. The spacing between the features depends on the rotational constants for the ground and excited states, as shown in the inset. We observe spectra by driving transitions from both $(0,5)_\mathrm{G}$ and $(1,6)_\mathrm{G}$. The green feature corresponds to the $\sigma^{-}$ transition $(1,6)_\mathrm{G} \rightarrow (0,5)_\mathrm{E}$ already shown in \fig{fig:vib_spectra}, and is recorded using light polarised perpendicular to the quantisation axis. 
 
\fig{fig:all_transitions}(c,d) show the hyperfine structure associated with the red and blue features in (a). The red features with solid markers were recorded using light polarised parallel to the quantisation axis, capable of driving $\pi$ transitions. The open red markers represent $\sigma^{\pm}$ transitions, recorded using light perpendicular to the quantisation axis. \fig{fig:all_transitions}(e)\,--\,(k) labels the individual components according to the excited-state hyperfine character. There are three $\pi$ transitions from $(0,5)_\mathrm{G}$ to states with character $(1,5)_\mathrm{E}$, as shown in \fig{fig:all_transitions}(e,\,h,\,i). In addition, we initially observed five $\sigma^\pm$ transitions, as shown in \fig{fig:all_transitions}(f,\,g,\,h,\,k). Of these, four are $\sigma^{-}$ transitions from $(0,5)_\mathrm{G}$ to states with character $(1,4)_\mathrm{E}$ and one is a $\sigma^{+}$ transition from $(0,5)_\mathrm{G}$ to the single stretched state with character $(1,6)_\mathrm{E}$. To identify the $\sigma^{+}$ transition, we used right-circularly polarised light and were able to drive only one of the five perpendicular transitions from $(0,5)_\mathrm{G}$. 

Theory predicts the existence of six hyperfine states with character $(1,4)_\mathrm{E}$. As described above, we initially observed only four of these. However,  after fitting our initial results to the theoretical model discussed in Sec.~\ref{sec:modelhfs}, we searched for the two missing transitions in the regions predicted by the model, shown by the open red markers in \fig{fig:all_transitions}(i,j); they are significantly weaker than the transitions observed initially.

The blue features in \fig{fig:all_transitions}(d) were recorded using light polarised parallel to the quantisation axis. They are transitions from $(1,6)_\mathrm{E}$ to states with character $(2,6)_\mathrm{E}$, as we drove only $\pi$ transitions. These three transitions are spread over approximately 12\,MHz. 

We have performed similar, though less thorough, spectroscopic measurements for transitions $v''\,= 0\,\rightarrow\,v' = 1,\, 2$. In $v' = 1$ we have mapped out all the excited states labelled by $(0,5)_\mathrm{E},(1,5)_\mathrm{E},(2,6)_\mathrm{E}$ and in $v' = 2$ we have measured the transition to $(0,5)_\mathrm{E}$. 

Our measurements were performed over a period of several months. To negate the effects of our reference cavity drifting (around $100$\,kHz per month) we reference our measurements to the transition $(1,6)_\mathrm{G} \rightarrow (0,5)_\mathrm{E}$ associated with that vibrationally excited state. These relative frequencies can be measured precisely using the known free-spectral range of the cavity and the applied frequency offset of the laser from the nearest cavity mode. The measured transition frequencies are given in \tab{tab:rot_hf_transitions}. We also give the \emph{hyperfine shifts}, defined as the transition frequencies relative to the lowest-energy component in that rotational manifold.

During our measurement campaign we observed that the transition to $(1,4)_\mathrm{E}$ shown in \fig{fig:all_transitions}(j) is subject to an intensity-dependent light shift. The shift is quadratic and shifts the frequency of the transition downwards as the light intensity increases. We believe this is caused by the light coupling the excited state to states labelled by $(0,3)_\mathrm{G}$; the nearest such state is detuned from~$(0,5)_\mathrm{G}$ by only 124\,kHz. This coupling is available only for $(1,4)_\mathrm{E}$ excited states; for the states $(1,5)_\mathrm{E}$ and $(1,6)_\mathrm{E}$ there is only one ground state with coupling allowed by the selection rules. To mitigate this effect we performed the spectroscopy on this line at a sufficiently low intensity that any residual light shift is negligible. There may be small light shifts for transitions to other states $(1,4)_\mathrm{E}$, but we estimate that they are smaller than the FWHM of the respective features and comparable to the statistical uncertainties in line positions. 

\renewcommand{\arraystretch}{1.25}
\begin{table*}[t!]
\caption{\label{tab:rot_hf_transitions} Summary of the spectroscopic measurements performed at 181.6\,G. The columns show: the vibrational levels involved in the transition ($v'' \rightarrow v' $); the ground state $(J'',M''_F)_{\rm{G}}$; the character of the transition; the hyperfine assignment of the excited state $(J',M'_F)_{\mathrm{E}}$; the relative transition frequency calculated with respect to the transition $(1,6)_\mathrm{G}\rightarrow (0,5)_\mathrm{E}$ for each vibrational level; the hyperfine shift calculated with respect to the lowest energy state of each rotational manifold; effective magnetic moment ($\mu_\mathrm{eff}$) of the excited state over a defined field range of $180$~G to $210$~G. The hyperfine shifts obtained from the theory model (\texttt{Fit 1}) are included for comparison.}
\centering
\begin{NiceTabular}{c|c|c|c|p{2.5cm}|c|p{2.5 cm}|p{2.5 cm}|c}
\hline \hline 
$v''\rightarrow v'$ & $(J'', M''_F)_\mathrm{G}$ & Transition  & $(J',M'_F)_\mathrm{E}$ & Transition frequency (MHz) & FWHM (MHz) & Hyperfine shift from experiment (MHz)& Hyperfine shift from \texttt{Fit1} (MHz) & $\mu_\mathrm{eff} / \mu_\textrm{N}$  \\ 
\hline

\Block{12-1}{$0\rightarrow 0$}      & $(1,6)_\mathrm{G}$    & $\sigma^{-}$   & $(0,5)_\mathrm{E}$        & $0$ & $0.046(3)$                     &   &            & $-18.6(1)$                  \\ \cline{2-9}

& \Block{8-1}{$(0,5)_\mathrm{G}$}   & $\pi$     & $(1,5)_\mathrm{E}$    & $2004.792(2)$& $0.039(4)$    & $0$   & $-0.21$ &$-26.7(4)$      \\
&   & $\sigma^{-}$   & $(1,4)_\mathrm{E}$    & $2005.491(4)$ & $0.058(11)$  & $0.700(5)$    &   $0.49$ &                   \\
&   & $\sigma^{-}$   & $(1,4)_\mathrm{E}$    & $2011.603(2)$ & $0.037(4)$   & $6.812(3)$    &  $6.83$ &                 \\
&   & $\sigma^{-}$   & $(1,4)_\mathrm{E}$    &  $2015.608(3)$ & $0.051(7)$   & $10.817(3)$   &  $11.51$ &                 \\
&   & $\sigma^{+}$   & $(1,6)_\mathrm{E}$    &  $2015.789(2)$ & $0.041(5)$  & $10.998(3)$   &  $10.94$ &                 \\
&   & $\pi$                     & $(1,5)_\mathrm{E}$    & $2015.943(1)$& $0.037(2)$   & $11.152(2)$   & $11.22$ & $-11.9(5)$      \\
&   & $\sigma^{-}$                     & $(1,4)_\mathrm{E}$    &  $2017.258(6)$& $0.051(16)$    & $12.467(6)$   & $12.87$ &       \\
&   & $\pi$                     & $(1,5)_\mathrm{E}$    &  $2019.345(2)$& $0.051(4)$   & $14.554(2)$   & $14.58$ & $-10.1(5)$      \\
&   & $\sigma^{-}$                     & $(1,4)_\mathrm{E}$    &  $2019.663(8)$& $0.079(19)$  & $14.871(8)$   & $14.82$ &       \\
&   & $\sigma^{-}$   & $(1,4)_\mathrm{E}$    &  $2022.587(1)$& $0.045(3)$   & $17.796(2)$ &    $17.75$ &               \\ \cline{2-9}

& \Block{3-1}{$(1,6)_\mathrm{G}$}    & $\pi$     & $(2,6)_\mathrm{E}$   &  $3091.352(1)$& $0.041(3)$   & $0$ & $0.04$   & $-20.2(2)$       \\
&   & $\pi$                     & $(2,6)_\mathrm{E}$    &  $3100.718(1)$& $0.045(2)$   & $9.366(2)$    & $9.31$ & $-9.8(3)$       \\
&   & $\pi$                     & $(2,6)_\mathrm{E}$    &  $3103.627(1)$ & $0.045(2)$  & $12.275(2)$   & $12.26$ & $-8.0(3)$       \\ \hline

\Block{8-1}{$0\rightarrow 1$}       & $(1,6)_\mathrm{G}$    & $\sigma^{-}$   & $(0,5)_\mathrm{E}$    &  $0$ & $0.037(4)$  &  &     & $-18.5(1)$          \\ \cline{2-9}

& \Block{3-1}{$(0,5)_\mathrm{G}$}   & $\pi$      & $(1,5)_\mathrm{E}$   & $2003.596(2)$ & $0.034(2)$   & $0$    &   & $-25.9(3)$      \\
&  & $\pi$      & $(1,5)_\mathrm{E}$   &  $2014.197(1)$ & $0.040(2)$  & $10.600(2)$ &  & $-13.7(2)$              \\
&  & $\pi$      & $(1,5)_\mathrm{E}$   &  $2014.465(2)$ & $0.034(3)$  & $10.869(2)$ &  & $-11.3(2)$              \\ \cline{2-9}

& \Block{3-1}{$(1,6)_\mathrm{G}$}     & $\pi$           & $(2,6)_\mathrm{E}$    &  $3087.249(2)$& $0.032(2)$ &  $0$    &   &  $-20.0(2)$       \\
&   & $\pi$     & $(2,6)_\mathrm{E}$  &  $3096.148(2)$& $0.033(3)$  & $8.899(2)$         &   & $-11.3(2)$      \\
&   & $\pi$     & $(2,6)_\mathrm{E}$  &  $3096.370(2)$& $0.034(3)$ & $9.121(2)$         &   &  $-9.6(2)$      \\ \hline

\Block{1-1}{$0\rightarrow 2$}  & $(1,6)_\mathrm{G}$     & $\sigma^{-}$ & $(0,5)_\mathrm{E}$&  $0$ &  $0.040(3)$   &    &  &  $-18.6(2)$ \\
\hline \hline
\end{NiceTabular}
\end{table*}

\subsection{Zeeman shift}\label{sec:Zeeman_shift}

\begin{table}[h!]
\centering
\caption{\label{tab:fit_parameters}%
Fitted values of the parameters for the linear and quadratic fits to the transition frequency of $(1,6)_\mathrm{G} \rightarrow (0,5)_\mathrm{E}$ as functions of magnetic field along with the reduced $\chi^2~(\chi^2_{\nu})$.}
    \begin{ruledtabular}
    \begin{tabular}{lcc}
    & Linear fit  & Quadratic fit \\
    & $(a_{0}+a_{1}x)$    & $(a_{0}+a_{1}x+a_{2}x^2)$ \\
    \hline
    $a_{0}$ (MHz)       & 1.84(2)        & 1.77(1)\\
    $a_{1}$ (MHz/G)     & $-0.0102(1)$ & $-0.0093(1)$\\
    $a_{2}$ (MHz/G$^2$) &     & $-0.0000022(3)$\\
    $\chi^2_{\nu}$ & $14.3$ & $1.6$\\
    \end{tabular}   
    \end{ruledtabular}   
\end{table}

To characterize the structure further, we measure the shifts in the transition frequencies as a function of the magnetic field. We accomplish this by recapturing the molecules in the dipole trap and ramping the magnetic field to the target value before turning the trap off again and driving the optical transition. The molecules are then recaptured again, the magnetic field is ramped back to $181.6$~G, and the remaining molecules are detected. For each transition, we measure the transitions in the region from 180~G to 210~G, where we observe a quasilinear Zeeman shift. We perform a linear fit to the measurements to extract the Zeeman shift of the transition. Then to extract the effective magnetic moment of the excited state in this magnetic field range $(\mu_\mathrm{eff})$ we subtract the well-known magnetic moment of the spin-stretched ground states ($-5.3$\,$\mu_\mathrm{N}$). The values of the effective magnetic moments are included in \tab{tab:rot_hf_transitions}.

\begin{figure}[t!]
    \centering
    \includegraphics[width=0.5\textwidth]{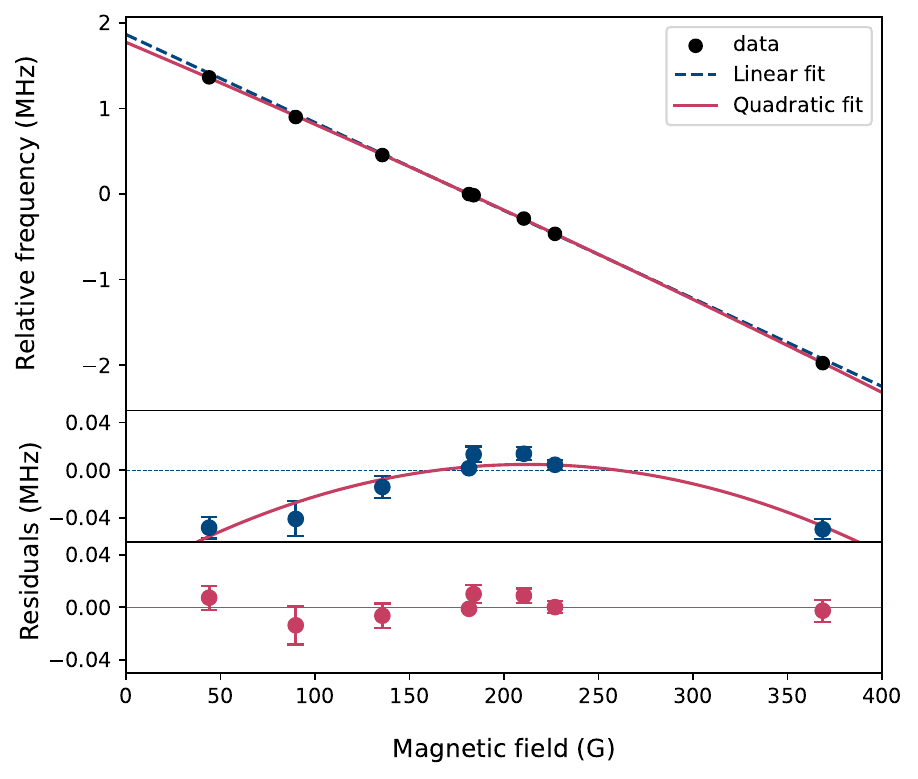}
    \caption{\label{fig:zeemanshift} Shift in frequency with magnetic field for the $(1,6)_\mathrm{G} \rightarrow (0,5)_\mathrm{E}$ transition between $44$ and $369$~G. To characterise the shift we fit the data (black markers in upper panels) with both linear (blue dashed line) and quadratic functions (red solid line). The residuals of both fit functions are shown in the bottom panel, where the blue and red markers correspond to the linear and quadratic fits, respectively. The red solid curve overlaid on the linear-fit residuals represents the difference between the quadratic and linear fitted values, indicating the quadratic nature of the shift.
    }
\end{figure}

We have performed more detailed measurements of the spin-stretched transition, $(1,6)_\mathrm{G} \rightarrow (0,5)_\mathrm{E}$. For this we measure the shift for magnetic fields from $44$~G to $369$~G, with respect to the frequency at $181.6$~G. The resulting shifts are shown in \fig{fig:zeemanshift}. We fit the shifts with linear and quadratic functions, giving the parameters in \tab{tab:fit_parameters}.   
It may be seen that the quadratic term is significant when we compare the reduced $\chi^2 (\chi^2_\nu)$ values between both fittings and also the residuals shown in the bottom panel of \fig{fig:zeemanshift}. The residuals of the linear fit deviate more in the small and large magnetic field regions than those of the quadratic fit, indicating that the observed shift includes a quadratic component. This is further supported by the difference between the quadratic and linear fitted data (red solid curve on the linear-fit residuals), which reproduces the same curvature as the linear-fit residuals. As will be seen in Sec.~\ref{sec:modelhfs}, the quadratic term provides valuable information on the position of the $0^-$ state that is otherwise hard to extract.

\subsection{Model for the rotational and hyperfine structure of the $v' = 0$ level  of the \btpi\, state}
\label{sec:modelhfs}

We model the substructure of the vibrational level $v' = 0$ of the $\mathrm{b}\,^3\Pi_{0^+}$ state of RbCs, for which the most experimental data are recorded here (\tab{tab:rot_hf_transitions}). We first define an effective Hamiltonian along the lines similar to~\cite{aldegunde2017}. Considering the scale of energies involved, we start with the zero-order Hamiltonian $\hat{H}_0$ with energies $E_{0^+}$ and $E_{0^-}$, corresponding to the $v'=0$ levels of these electronic states. We treat the rotational Hamiltonian $\hat{H}_\mathrm{r}$, the hyperfine Hamiltonian $\hat{H}_\mathrm{hf}$, and the Zeeman Hamiltonian $\hat{H}_\mathrm{Z}$ as perturbations of $E_{0^+}$. Our effective Hamiltonian $\hat{H}_{\textrm{eff}}$ is
\begin{equation}
  \hat{H}_{\textrm{eff}} = \hat{H}_0 + \hat{H}_\mathrm{r} + \hat{H}_\mathrm{hf} + \hat{H}_\mathrm{Z},
\end{equation}
with
\begin{align}
  \hat{H}_\mathrm{r} &= B_v \vec{J}^2 ,\\
  \hat{H}_\mathrm{hf} &= \sum_{i=\textrm{Rb,Cs}} e\bar{Q}(i) . \bar{q}(i)  + \sum_{i=\textrm{Rb,Cs}} \hat{H}_{\textrm{MD}}(i) ,\\
  \hat{H}_\mathrm{Z} &= g_J \mu_\textrm{B} \vec{J} . \vec{B} + \sum_{i=\textrm{Rb,Cs}} g_i(1-\sigma_i)\mu_\textrm{N} \vec{I}(i) . \vec{B} \,.
\end{align}
The rotational part $\hat{H}_\mathrm{r}$ involves the rotational constant $B_v$ of the level concerned (here the level $v=0$ of the $0^+$ component), with $\vec{J}$ the total angular momentum of the molecule excluding nuclear spins.  The hyperfine Hamiltonian $\hat{H}_\mathrm{hf}$ contains several terms. The first is the interaction between the electric quadrupole moment tensors of the nuclei $e\bar{Q}(i)$ and the electric-field-gradient tensors $\bar{q}(i)$ due to the electrons, with coupling constants $(eQq)_{\textrm{Rb}}$ and $(eQq)_{\textrm{Cs}}$. The second term, involving $\hat{H}_{\textrm{MD}}$(Rb) and $\hat{H}_{\textrm{MD}}$(Cs), accounts for magnetic-dipole interactions between the nuclear magnetic moments and those due to electron spin, orbital angular momentum and rotation. These terms all have the same angular structure~\cite{broyer1978}, so they are characterized by two effective coupling constants $A_{\textrm{Rb}}$ and $A_{\textrm{Cs}}$, as described the Appendix. The interaction between the two nuclear spins is neglected, as it is expected to be very small~\cite{aldegunde2008}. The term $\hat{H}_\mathrm{Z}$ includes the interactions of magnetic field $\vec{B}$ with $\vec{J}$, described by an effective $g$-factor $g_J$, and with nuclear spins, involving the nuclear $g$-factor $g_i$ and the shielding factor $\sigma_i$ for each nucleus $i$. The expressions for the corresponding matrix elements are given in the Appendix. 

The Hamiltonian $\hat{H}_{\textrm{eff}}$ is expressed on the basis $\ket{nJM_J p, I_{\textrm{Rb}}m_\mathrm{Rb}I_{\textrm{Cs}}m_\mathrm{Cs}}$, with $n$ corresponding to the vibronic part (one vibrational level for each of $0^+$ and $0^-$), including $J=0,1,2,3$, and all possible projections that satisfy $M_F = M_J + m_\mathrm{Rb} + m_\mathrm{Cs}$. The total parity, $p=(-1)^J$ for $0^+$ and $(-1)^{J+1}$ for $0^-$, is a conserved quantity that reduces the size of the basis. The Hamiltonian is diagonalized to obtain eigenvalues and associated eigenvectors that model the spectrum.

$\hat{H}_{\textrm{MD}}(i)$ and the first term in $\hat{H}_\mathrm{Z}$ impose the selection rule $\Delta J = \pm 1$ for each parity. They do not have diagonal matrix elements for $\Omega=0$; only off-diagonal elements between $0^+$ and $0^-$ contribute to the hyperfine structure. These two operators are the dominant terms in the hyperfine structure of alkali-metal atoms and can induce strong effects in a dimer. However, if the energy gap $\Delta$ between $0^+$ and $0^-$ is large enough, they produce only weak perturbations. This explains the small hyperfine splitting observed for $\Omega=0^\pm$, compared to those for $\Omega=1$, where the hyperfine splittings are in the GHz range~\cite{orban2015,orban2019}.

To ensure the best possible least-squares fitting of the experimental data and to reduce correlations between the fitted parameters, it is desirable to reduce the number of free parameters. We note that, for all operators, off-diagonal couplings between two different rotational levels $J$ are either very small or represent a perturbation between the $0^+$ and $0^-$ states. If the rotational constants of these two levels are much smaller than their energy separation $\Delta$, their influence on the observed hyperfine shift will be minor. Thus, before starting our fitting procedure, we fixed the rotational constant $B_0(0^+) \approx 510$ MHz, based on the spectroscopic study of Docenko et al.~\cite{Docenko2010}. We checked after the fitting procedure that this approach was accurate, as significant variations in the value of $B_0$ do not affect the fit results.

For the electronic ground state, $(eQq)_{\textrm{Rb}} \gg (eQq)_{\textrm{Cs}}$~\cite{aldegunde2008, Gregory2016}. In the present work, there are insufficient experimental data to determine the corresponding quantities for b$^3\Pi_{0^+}$ for both nuclei, so we set $(eQq)_\textrm{Cs}=0$. The quantity $(eQq)_{\textrm{Rb}}$ is retained as a fitting parameter. It depends on the gradient of the electric field due to the electrons, so is different in the ground and excited states. The nuclear Zeeman coupling constants are set equal to the ground-state values $g_{\textrm{Rb}}(1-\sigma_{\textrm{Rb}})=1.8295$~\cite{Gregory2016} and $g_{\textrm{Cs}}(1-\sigma_{\textrm{Cs}})=0.7331$~\cite{Gregory2016}. 

The matrix elements of the non-nuclear part of $\hat{H}_\mathrm{Z}$ follow the selection rule $\Delta J=\pm1$ for $\Omega=0$, and are therefore all off-diagonal between $0^+$ and $0^-$. They are much smaller than the separation $\Delta$ between $0^+$ and $0^-$ and are proportional to $B$, so they produce a contribution to the Zeeman effect that is quadratic in $B$. The magnitude of their matrix elements is known \emph{a priori}, so that the quadratic Zeeman effect described in \tab{tab:fit_parameters} provides direct information on $\Delta$, which is otherwise difficult to determine.

Once the parameters of the model are set, we need to compare the computed energies and magnetic shifts to the experimental ones to obtain residuals that will be used in the least-squares fit. Because hyperfine shifts are largely insensitive to the rotational constant, we computed the residuals of the manifolds for $J' = 1$ and $J' = 2$ independently. However, the origin of energy used to present the experimental data -- chosen for each rotational manifold as the position of the lowest-frequency line -- does not match the origin of energy of the model, which is the energy of the spin-free level ($0^+,\,v' =0,\,J'$). Therefore, two additional free parameters $E_{J' = 1}$ and $E_{J' = 2}$ are needed to align the calculated hyperfine structure with the experimental measurements. These correspond to the energy difference between centre of gravity of the rotational manifold and the lowest-energy hyperfine state experimentally observed in that rotational manifold.

In summary, there are six free parameters of the model. These are the energy gap $\Delta$ between the lowest rovibrational levels of $0^+$ and $0^-$, the magnetic dipolar coupling constants $A_{\textrm{Rb}}$ and $A_{\textrm{Cs}}$, the electric quadrupole coupling constant $(eQq)_{\textrm{Rb}}$, and the energy shifts $E_{J' = 1}$ and $E_{J' = 2}$. 

The least-squares fit is performed using the Python library \texttt{lmfit} with the Levenberg-Marquardt algorithm. The initial set of experimental data included in the fit are the thirteen energies of the $J' = 1$ and $J' = 2$ rotational levels of $v' = 0$ and the six related effective magnetic moments; see \tab{tab:rot_hf_transitions}. In addition, the quadratic fit to the Zeeman shift for the transition $(1,6)_\mathrm{G} \rightarrow (0,5)_\mathrm{E}$, described above (\tab{tab:fit_parameters}), produces three magnetic data for $J' = 0$, namely the parameters $a_0$, $a_1$ and $a_2$. This gives a total of 22 data points to fit.

Despite the simplifications described above, the fit was unstable, due to a strong correlation between the parameters. To resolve this, we began with a sequential fit (referred to as \texttt{Fit1} in the following) that takes advantage of physical insight into the ways in which particular features of the spectra depend on the parameters of the model.

As noted above, the quadratic coefficient of the Zeeman effect in \tab{tab:fit_parameters} depends mainly on the energy gap $\Delta$ between the $v'=0$ levels of the $0^+$ and $0^-$ states. Thus, we run the fitting procedure with only five free parameters and with $\Delta$ frozen to an arbitrary value. This procedure is repeated for many values of $\Delta$, allowing us to identify sets of parameters that correctly reproduce the quadratic Zeeman shift. Since the simplified Hamiltonian of the model ignores several small contributions, the fit is not expected to reach the experimental accuracy (estimated at a few kHz). The resulting parameters are reported in \tab{tab:model_parameters}, and the experimental energy levels are reproduced with a mean error of 240~kHz. The uncertainty on $\Delta$ is obtained first, from the variation of the quadratic Zeeman coefficient $a_2$ (\tab{tab:fit_parameters}) with $\Delta$: the experimental uncertainty on $a_2$ determines the uncertainty on $\Delta$. The uncertainty on the other parameters is the quadratic sum of two independent uncertainties of comparable magnitude. The first is provided by the fitting routine itself. The second is deduced from a Monte-Carlo exploration of the variation of the parameters for about 1000 values of $\Delta$ randomly chosen within its uncertainty interval.

\begin{figure}[t!]
    \centering
    \includegraphics[width=0.5\textwidth]{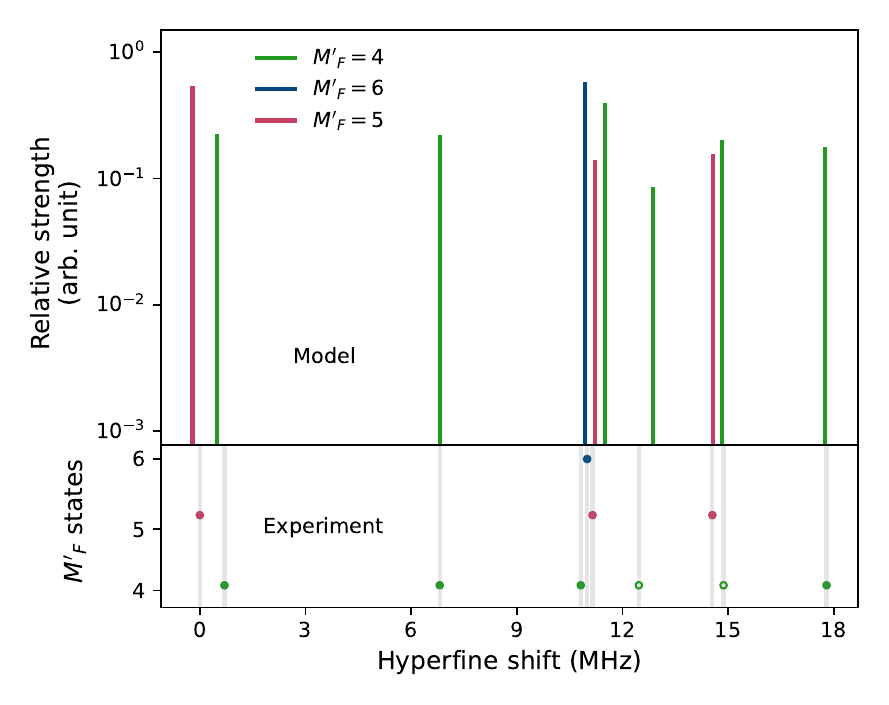}
    \caption{Comparison between experimental and theoretical relative transition frequencies (hyperfine shifts) from $J''=0$ to $J'=1$. The upper panel shows the relative transition frequencies and the relative transition strengths extracted from the model (shown in \tab{tab:rot_hf_transitions}), while the lower panel shows the experimentally observed relative frequencies only. The x-axis shows the transition frequency relative to the observed lowest-energy state with $M'_F = 5$. Solid lines in the upper panel are shown in green, red and blue, corresponding to upper states with $M'_F = 4,\, 5$ and $6$, respectively. The experimental observations are shown with markers in the lower panel using the same colours. The vertical position of the marker indicates the $M'_F$ value, as shown on the y-axis. Grey vertical lines in the lower panel are included as visual guides to facilitate comparison between the experimental and theoretical results. The transitions that were initially unobserved in the experiment are indicated by open green markers.}
    \label{fig:stick_spectra}
\end{figure}

Only four of the six lines predicted to states with $J'=1,\, M'_F=4$ were initially recorded, making their assignment unclear. We ignored them in the first step. After obtaining reliable values for the parameters, the model was precise enough to assign the four recorded lines, allowing us to refine the fit by including them and to predict the position of the two missing lines. The initially missing lines were then measured, as described in Sec.~\ref{sec:rot_hf_structure}, and included in the final fit. To compare experiment and theory, we show the relative transition frequencies in the form of a stick spectrum in \fig{fig:stick_spectra}. 

The fit yields the positions of the centre of gravity of the manifolds for $J'=0$, 1 and 2. This allows us to model the rotational energies proportional to $J'(J'+1)$. We extract $B_0(0^+) = 515.9(1)$~MHz, with an uncertainty related to those on $E_{J'=1}$ and $E_{J'=2}$. Our result agrees with the value $B_0 = 516$~MHz calculated from the potential curves of Ref.~\cite{Kruzins2014}, but is more precise because hyperfine structure was not resolved in the earlier work.

\tab{tab:rot_hf_transitions} shows that \texttt{Fit1} yields line positions with an average accuracy of about 240~kHz, which is much larger than the  experimental uncertainty. This is mostly due to the neglected terms in the model Hamiltonian. Nevertheless, to assess the robustness of the results, we relaxed the constraints of \texttt{Fit1} by treating $B_0$ and $\Delta$ as free parameters, resulting in a new fit \texttt{Fit2}. The rotational manifolds are now included simultaneously, using the transition frequencies of \tab{tab:rot_hf_transitions} (instead of the hyperfine shifts with respect to the center of gravity), so we replace the parameters $E_{J'=1}$ and $E_{J'=2}$ by a single global shift named $E_{J'=0}$. To ensure convergence when calculating the residuals, we replace the initial option in \texttt{lmfit} to evaluate the uncertainties on energies by another option that imposes a fixed model uncertainty $\sigma_{\textrm{model}}$ in addition to the experimental uncertainty. The fitted parameters of \texttt{Fit1} are the starting values for \texttt{Fit2}. The optimal fit uses $\sigma_{\textrm{model}}=160$~kHz and results in a value of $\chi^2_\textrm{red}$ close to 1, with an average accuracy of 179~kHz in the line positions. We see in \tab{tab:model_parameters} that the fitted values of $\Delta$ and $B_0$ are robust, as the two fits are consistent with each other. The parameter that changes the most between the two fits is $(eQq)_{\textrm{Rb}}$, which characterises the weakest interaction in the model; this suggests that its value partially compensates for the limitations of the model.

\begin{table}[h!]
\centering
\caption{\label{tab:model_parameters} Best fitted values (with uncertainties) of the parameters involved in the model for the substructure of the $v=0$ level of the $0^+$ state.}
     \begin{ruledtabular}
     \begin{tabular}{lcccc}
     &\multicolumn{2}{c} {\texttt{Fit1}}& \multicolumn{2}{c}{\texttt{Fit2}} \\
 Parameter    & Value  & Uncertainty & Value  & Uncertainty \\
     \hline
     $\Delta$ (GHz)       & $291$       & $20$ & $294$& $20$\\
     $A_{\textrm{Rb}}$ (MHz) & $940$ & $92$ & $987$ & $86$\\
     $A_{\textrm{Cs}}$ (MHz) &   $379$  & $37$ & $368$& $32$ \\
     $(eQq)_{\textrm{Rb}}$ (MHz) & $9.7$ & $0.8$ & $12$& $1$\\
    $E_{J'=0}$ (MHz) & - & - & $13.4$ & $0.4$ \\
    $E_{J'=1}$ (MHz) & $20.76$& $0.34$ & - &-\\
    $E_{J'=2}$ (MHz) & $17.53$ & $0.37$ & - &-\\
    $B_0$ (MHz) & $515.96$ & $0.10$ & $515.95$ & $0.03$\\
     \end{tabular}
     \end{ruledtabular}
\end{table}

For completeness, we tried to refine the fit by treating $(eQq)_\textrm{Cs}$ and the nuclear $g$ factors as free parameters. This did not give reliable values for these parameters due to their expected weak contributions.

\subsection{Transition dipole moments}
\label{sec:TDM}

We measure the vibronic transition dipole moment (TDM) by driving Rabi oscillations on the spin-stretched transitions $(0,5)_\mathrm{G} \rightarrow (1,6)_\mathrm{E}$. The Rabi frequency here is related to the TDM through
\begin{equation} \label{eq:Rabi and TDM}
    \hbar\Omega=\frac{1}{\sqrt{3}}\mu_{0, v'} E = \frac{1}{\sqrt{3}}\mu_{0, v'}\sqrt{\frac{2I}{c \epsilon_0}},
\end{equation}
where $\Omega$ is the angular Rabi frequency, $\mu_{0, v'}$ is the TDM, $E$ is the amplitude of the electric field of the probe light driving the oscillation with intensity $I$. The transition linewidth $\Gamma_{0, v'}$ is then related to the TDM by
\begin{equation} \label{eq:transition linewidth and TDM hyperfine}
    \Gamma_{0, v'}=\frac{\omega_{v'}^3}{3\pi\epsilon_0\hbar c^3}|\mu_{0, v'}|^2,
\end{equation}
where $\omega_{v'}$ is the angular transition frequency.

\begin{figure}[t!]
    \centering
    \includegraphics[width=0.48\textwidth]{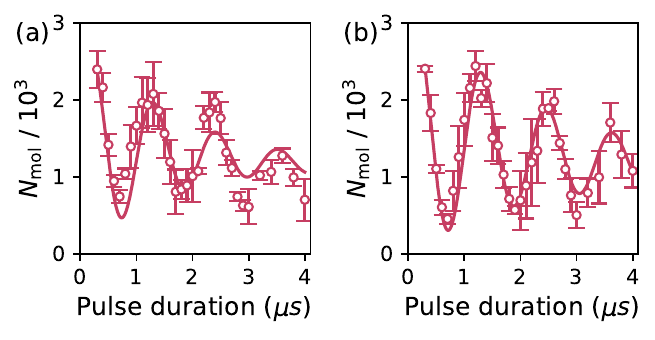}
    \caption{\label{fig:Rabi oscillations}
    Rabi oscillations between $(0,5)_\mathrm{G}$ and $(1,6)_\mathrm{E}$ for (a) $v' = 0$ (b) $v' = 1$. The lines are fits of a damped Rabi oscillation model to the data. Note the two measurements were taken using different probe intensities $30(1)\, \mathrm{mW}/\mathrm{cm}^2$ and $46(2)\, \mathrm{mW}/\mathrm{cm}^2$ respectively.}
\end{figure}

We measure Rabi oscillations for the transitions $(0,5)_\mathrm{G} \rightarrow (1,6)_\mathrm{E}$ for $v' = 0,\, 1$ using circularly polarised light propagating approximately along the quantisation axis. We pulse the resonant probe light on with an acousto-optic modulator for a variable duration and then measure the number of ground-state molecules remaining as a function of the pulse time. The results are shown in \fig{fig:Rabi oscillations}. We fit the data with a damped oscillation to extract the Rabi frequencies. The calculated values of TDM and transition linewidth from the experimental measurement are compared with the theoretical values in \tab{tab:linewidths}.

There are two sources of uncertainty in the TDMs and the transition linewidths. The first is the uncertainty of the Rabi frequencies from fitting and the second is the shot-to-shot fluctuation in the probe power. For the latter, we record the power for each shot and then compute the standard deviation over the measurement to quantify the uncertainty.

\begin{table}[t]
\centering
\begin{ruledtabular}
   \begin{tabular}{c|cccc|cc}
        $v'$ & \multicolumn{1}{c}{$\Omega_{0, v'}^\mathrm{hf}$} & \multicolumn{1}{c}{$\mu_{0, v'}$} & \multicolumn{2}{c|}{$\Gamma_{0, v'}$ (kHz)} & $\tau_0$ & \multicolumn{1}{c}{$\gamma_{v}^\mathrm{e}$}\\
         & (kHz) & (Debye) & Theory~\cite{Guan2021} & This work & ($\mu$s) & (kHz) \\
         \hline
         0& $897(7)$ & $0.65(2)$ & $15.5$ & $14.1(7)$ & $12.3(1)$ & $13.0(9)$ \\
         1& $855(7)$ & $0.50(1)$ & $6.84$ & $8.2(4)$ & $7.20(4)$ & $22.1(8)$  \\
           
    \end{tabular} 
\end{ruledtabular}
\caption{The TDM and linewidths\footnote{\label{1}
The values of $\Gamma_{0,v'}, \gamma_{v'}^\mathrm{e}$ are reported in Hz (cycles/s). Conversion to angular units (rad/s) requires multiplication of the reported numerical values by $2\pi$~\cite{Mohr2015}.} 
of $v'=~0$ and $v'=~1$ of the \btpi\, state. The TDM and transition linewidths are calculated from the Rabi frequencies of \fig{fig:Rabi oscillations} and the associated intensities. The lifetime and natural linewidth of the excited states are derived from the loss measurement of \fig{fig:loss rate}. 
}
\label{tab:linewidths}
\end{table}

We observe that there is a decay in the oscillation amplitude due to dephasing. The decay times of the Rabi oscillations amplitudes for $v' = 0$ and $1$ are also extracted from fitting as $1.8(1)\, \mu$s and $3.1(2)\, \mu$s respectively. We attribute the dephasing to a combination of factors: laser power fluctuations (both from shot to shot and during a single pulse), frequency noise on the laser due to noise on the PDH lock, magnetic field noise which affects the transition frequency, Doppler shifts due to molecular motion and small inhomogeneities in the intensity distribution over the molecule cloud.

\begin{figure}[t!]
    \centering
    \includegraphics[width=0.48\textwidth]{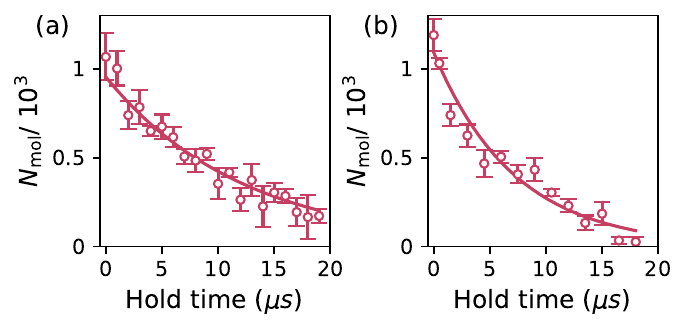}
    \caption{Measurement of the lifetimes of molecules prepared in the state $(1,6)_\mathrm{E}$ for (a) $v' = 0$ and (b) $v' = 1$.}
    \label{fig:loss rate}
\end{figure}

\subsection{\label{sec:linewidth} Natural linewidths of the excited state}

The vibrational levels of the \btpi\, state can decay on multiple transitions. This sets the lifetime of the excited state, $\tau_0$ and hence the natural linewidth, $\gamma^\mathrm{e}_{0, v'}=1/\tau_0$.

We measure the lifetime of the excited state $(1,6)_\mathrm{E}$ for $v'=0$ and $v'=1$ by directly measuring loss from this state. We transfer the molecules from $(0,5)_\mathrm{G}$ to $(1,6)_\mathrm{E}$ using a $\pi$ pulse of the resonant probe beam. The molecules are then held in the excited state for a variable time before they are transferred back to the ground state by another $\pi$ pulse of the probe. Around half of the molecules are lost due to imperfections in the $\pi$ pulses. The number of molecules in the ground state is measured as a function of the hold time, from which we extract the lifetime of the excited state by fitting to an exponential model of the decay. Loss measurements of $(1,6)_\mathrm{E}$ states for $v' = 0$ and at $v' = 1$ are shown in \fig{fig:loss rate}, and the lifetimes and hence the natural linewidths are given in \tab{tab:linewidths}. The uncertainties in their values are purely from the uncertainty of the fitting.

Our results are consistent with the transition to $v' = 0$ being vibrationally closed as $\Gamma_{0,0} \approx \gamma^\mathrm{e}_{0}$. The lower limit for the probability of decay from $v' = 0$ to $v'' = 0$ is greater than $79$\% at the 95\% confidence level. This is a significant difference from that observed for the equivalent transition in NaRb, where measurements suggested significant loss due to spontaneous emission to the triplet ground state $\mathrm{a}\,^3\Sigma^+$~\cite{He2021}.

\section{\label{sec:discussion} Conclusion}
We have investigated transitions from the lowest vibrational level of the electronic ground state \xssigma\, to the lowest few vibrational levels of \btpi\, and have observed well-resolved hyperfine structure associated with these transitions. We have measured vibrational, rotational and hyperfine splittings, effective magnetic moments, transition dipoles and excited-state linewidths. Our findings suggest that the $v'=0$ level has a greater than 79\% probability to decay to $v''=0$ at the 95\% confidence level, which make this transition a promising candidate for implementing proposed techniques for direct detection of bialkali molecules~\cite{Wang2010,Guan2020}. The narrow linewidth may also be useful for laser cooling~\cite{Shuman2010, Kobayashi2014}, as the associated Doppler temperature ($\sim 300$~nK) is significantly lower than the temperature currently achieved for gases of RbCs.

We have fitted the observed transition frequencies to obtain excited-state spectroscopic parameters. These include the rotational constant, hyperfine coupling constants, and the separation between the $0^+$ and $0^-$ components of the \btpi\, electronic state. This work complements previous spectroscopy performed on RbCs in heat pipes. The results will help us to develop an improved model of magic-wavelength trapping conditions and extend coherence times for ultracold RbCs molecules~\cite{Gregory2024,Hepworth2025}.

\section{Data Availability}
The data that support the findings of this article are openly available from Durham University~\cite{dataset}.

\section{Acknowledgements}
We acknowledge Svetlana Kotochigova for useful discussion and for providing the theoretical transition frequency for different vibrational levels in the excited state. We acknowledge support from the UK Engineering and Physical Sciences Research Council (EPSRC) Grants EP/P01058X/1, EP/W00299X/1 and UKRI2226 funded through the Programme Grant Scheme, UK Research and Innovation (UKRI) Frontier Research Grant EP/X023354/1, the Royal Society, and Durham University. PDG is supported by a Royal Society University Research Fellowship URF/R1/231274 and Royal Society research grant RG/R1/241149. AD is supported by Royal society Newton International Fellowship NIF/R1/232150. 

\bibliography{reference}

\appendix

\section{Matrix elements of the terms composing $\hat{H}_{\textrm{eff}}$}

Following~\cite{broyer1978,Cook1971}, the matrix elements are expressed in the uncoupled basis
\begin{align}
  \ket{i} =  \ket{n,JM_Jp,I_\textrm{Rb}m_\mathrm{Rb}I_\textrm{Cs}m_\mathrm{Cs}},\\
  \ket{j} = \ket{n,J'M'_Jp',I_\textrm{Rb}m'_\mathrm{Rb}I_\textrm{Cs}m'_\mathrm{Cs}}.
\end{align}
They are displayed below for Rb for simplicity, and similar terms are present for Cs.

The electric quadrupole operator, $\hat{H}_Q(\textrm{Rb})=\bar{V}(\mathrm{Rb}) \cdot \bar{Q}(\mathrm{Rb})$, has matrix elements
\begin{multline}
  \bra{j} H_{Q} \ket{i} = \delta_{pp'}\frac{(eQq)_{\textrm{Rb}}}{4} (-1)^{J'+J-\Omega'+I_{\textrm{Rb}}-M_J-m'_\mathrm{Rb}} \\ \sqrt{(2J+1)(2J'+1)} \tj{J'}{2}{J}{\Omega'}{\Delta\Omega}{\Omega}\tj{J'}{2}{J}{-M'_J}{\Delta M_J}{M_J} \\ 
  \sqrt{\frac{(2I_{\textrm{Rb}}+1)(2I_{\textrm{Rb}}+2)(2I_{\textrm{Rb}}+3)}{2I_{\textrm{Rb}}(2I_{\textrm{Rb}}-1)}} \tj{I_{\textrm{Rb}}}{2}{I_{\textrm{Rb}}}{-m'_\mathrm{Rb}}{\Delta m_\mathrm{Rb}}{m_\mathrm{Rb}}.
\end{multline}

The dipolar magnetic operator can be expressed in terms of a single effective coupling constant, along the lines summarized in Table II of~\cite{broyer1978}. It has matrix elements
\begin{multline}
  \bra{j} \hat{H}_{\textrm{MD}}(\textrm{Rb}) \ket{i} = \delta_{pp'}A_{\textrm{Rb}}(-1)^{J'+J-\Omega'+I_{\textrm{Rb}}-M_J-m'_\mathrm{Rb}} \\ \sqrt{(2J+1)(2J'+1)} \tj{J'}{1}{J}{\Omega'}{\Delta\Omega}{\Omega}\tj{J'}{1}{J}{-M'_J}{\Delta M_J}{M_J} \\ 
  \sqrt{I_{\textrm{Rb}}(I_{\textrm{Rb}}+1)(2I_{\textrm{Rb}}+1)} \tj{I_{\textrm{Rb}}}{1}{I_{\textrm{Rb}}}{-m'_\mathrm{Rb}}{\Delta m_\mathrm{Rb}}{m_\mathrm{Rb}}.
\end{multline}

The electronic term $\hat{H}_{\mathrm{Z}J}=g_J \mu_\textrm{B} \vec{J} \cdot \vec{B}$ in the Zeeman hamiltonian has matrix elements
\goodbreak
\begin{multline}
\bra{j} \hat{H}_\mathrm{Z} \ket{i} = \delta_{pp'}\delta_{M_JM'_J}\delta_{m_\mathrm{Rb}m'_\mathrm{Rb}}\delta_{m_\mathrm{Cs}m'_\mathrm{Cs}}\delta_{\Omega \Omega'} \\
\times (-(g_L-g_S)) \mu_\textrm{B} B_Z (-1)^{M_J-\Omega'}\sqrt{(2J+1)(2J'+1)} \\
\tj{J'}{1}{J}{\Omega'}{\Delta\Omega}{\Omega}\tj{J'}{1}{J}{M'_J}{\Delta M_J}{M_J},
\end{multline}
where $B_Z$ is the magnitude of the applied magnetic field.

The nuclear term $\hat{H}_{\mathrm{Z}I}$ (that is, the second term in $\hat{H}_\mathrm{Z}$) has matrix elements 
\begin{equation}
  \bra{j} \hat{H}_{\mathrm{Z}I} \ket{i} = -\delta_{ij} g_{\textrm{Rb}}(1-\sigma_{\textrm{Rb}}) \mu_\textrm{N} B_Z m_\mathrm{Rb}.
\end{equation}

\end{document}